\documentclass[english,aps,manuscript]{revtex4}
\usepackage[T1]{fontenc}
\usepackage[latin1]{inputenc}
\usepackage{amsmath}
\usepackage{color}
\usepackage{amssymb}

\makeatletter

\usepackage{color}

\makeatletter
\newcommand{\bee}{\begin{equation}}
\newcommand{\ee}{\end{equation}}
\newcommand{\beea}{\begin{eqnarray}}
\newcommand{\eea}{\end{eqnarray}}

\makeatother

\usepackage{babel}
\makeatother

\begin{document}

\section*{COLO-HEP-535}

\begin{center}
\textbf{\Large Mediation of Supersymmetry Breaking in a Class of String
Theory Models}{\Large{} }
\par\end{center}{\Large \par}

\begin{center}
\vspace{0.3cm}
 
\par\end{center}

\begin{center}
{\large S. P. de Alwis$^{\dagger}$ } 
\par\end{center}

\begin{center}
Physics Department, University of Colorado, \\
 Boulder, CO 80309 USA 
\par\end{center}

\begin{center}
\vspace{0.3cm}
 
\par\end{center}

\begin{center}
\textbf{Abstract} 
\par\end{center}

\begin{center}
\vspace{0.3cm}
 
\par\end{center}

A consistent theory of supersymmetry breaking must have a hidden sector,
an observable sector, and must be embedded in a locally supersymmetric
theory which arises from string theory. For phenomenological reasons
it must also transmit supersymmetry from the hidden to the visible
sector in a dominantly flavor neutral manner. Also any such theory
of supersymmetry breaking has to take into account the problem of
quadratic divergences which arise once the theory is embedded in supergravity.
A class of possible models that arise from GKP-KKLT type IIB string
compactifications, incorporating all this while being consistent with
gauge unification, with just the bare minimum of necessary supergravity/string
theory moduli fields coupled to the minimally supersymmetric standard
model, is presented. Such models give reasonable values for the soft
masses, the $\mu$ and $B\mu$ terms and the gaugino masses. Assuming
that an actual detailed realization exists, it is very likely that
they are the simplest such possibility .

\vfill{}

$^{\dagger}$ {\small e-mail: dealwiss@colorado.edu}{\small \par}

\eject

\section{Introduction\label{sec:Introduction}}

Much of the discussion of supersymmetry breaking in the Minimal Supersymmetric
Standard Model (MSSM) (and its generalizations) has been in the context
of global supersymmetry. However it is well-known that in order to
have a zero (or nearly zero) cosmological constant it is necessary
to incorporate supergravity (SUGRA) effects. This is usually done
by introducing a constant into the superpotential. The supergravity
potential, unlike the globally supersymmetric one, is not positive
definite and one can in principle use the constant to tune the cosmological
constant (CC) to zero. But once it is admitted that a consistent theory
needs to bring in supergravity effects, one needs to account for the
potential effects of quadratic sensitivity to high scale physics of
the low energy supersymmetry breaking parameters. Furthermore one
has to consider effects of the additional fields (moduli) that are
neutral under the standard model group but are essential ingredients
in any consistent supergravity such as string theory %
\footnote{For reviews of the MSSM, SUSY breaking mechanisms and phenomenological
SUGRA, see for example \citep{Weinberg:2000cr,Drees:2004jm,Baer:2006rs}.%
}.

The moduli that occur in any string theory construction need to be
stabilized, and in the recent literature there has been much discussion
of how this may be done, particularly in the context of type IIB string
theory %
\footnote{For reviews see \citep{Grana:2005jc,Douglas:2006es}.%
}. In general one can find minima for the moduli sector potential which
break supersymmetry. In fact generic minima would be expected to have
supersymmetry breaking at the natural scale of the theory - namely
the string scale. Nevertheless it is possible to find non-generic
points in this landscape which have a low or intermediate scale of
supersymmetry breaking. If supersymmetry is to be relevant for phenomenology,
the starting point of any string theory construction would have to
be one of these points. In the following we will work with type IIB
theory since in many respects this is the best understood, and we
assume that such points exist with the MSSM living on some brane configuration.
However we expect that similar arguments can be made in other string
theory contexts, and we suspect that the phenomenology (assuming the
relevant existence theorems can be established) is not likely to be
very different since our arguments rest on some general features of
the string theory input such as the tendency to have a `no-scale'
starting point at the classical level.

We work with $\kappa^{2}=8\pi G_{N}=M_{P}^{-2}=1$. A general supergravity
theory has a real analytic Kaehler potential $K=K(\Phi,\bar{\Phi})$
and a holomorphic superpotential $W(\Phi)$ where $\Phi=\{\Phi^{A}\},\, A=1,\ldots,N$,
is the set of chiral scalar fields of the theory. The metric on field
space is $K_{A\bar{B}}=\partial_{A}\partial_{\bar{B}}K$ and the metric
on the gauge group (which is in general a chiral function of the neutral
chiral superfields) is $f_{ab}(\Phi)$.

The embedding of a supersymmetry breaking theory in supergravity brings
in additional effects that are not usually considered in the literature.
The coefficient of the term that is quadratic in the cutoff in the
one loop effective potential, is proportional to ${\rm Str}M^{2}(\Phi)\equiv\sum_{J}(-1)^{2J}(2J+1){\rm tr}M^{2}(\Phi)$,
where $\Phi$ is the set of chiral (super) fields in the theory and
$M^{2}$ is the field dependent mass squared matrix. In a globally
supersymmetric theory (even if the supersymmetry is spontaneously
broken) this supertrace is zero and one has no quadratic divergence
in the quantum theory. However in a SUGRA theory whose supersymmetry
is spontaneously broken this supertrace does not vanish. Instead we
have %
\footnote{See for example \citep{Wess:1992cp}, \citep{Ferrara:1994kg}.%
}\begin{equation}
{\rm Str}M^{2}(\Phi)=(N-1)m_{3/2}^{2}(\Phi)-F^{A}(\Phi)(R_{A\bar{B}}+F_{A\bar{B}})(\Phi,\bar{\Phi})\bar{F}^{\bar{B}}(\bar{\Phi}),\label{eq:Str}\end{equation}
 where \begin{equation}
R_{A\bar{B}}=\partial_{A}\partial_{\bar{B}}\ln\det K_{C\bar{D}},\, F_{A\bar{B}}=-\partial_{A}\partial_{\bar{B}}\ln\det\Re f_{ab},\label{eq:HABbar}\end{equation}
 and $F^{A}$ is the F-term of the chiral multiplet $\Phi^{A}$ and
$m_{3/2}^{2}(\Phi)=e^{K}|W|^{2}$ is the field dependent gravitino
mass. 

Let the indices $I,J,\ldots$ denote fields of the visible (MSSM)
sector and $i,j,\ldots$ denote (closed string) moduli fields. Typically
they are expected to get vaccum expectation values (vevs) of the order
of the Planck scale or larger. Let the total number of such chiral
superfields in the visible sector be $N_{v}$. Note that this number
is taken to include GUT fields if any. In gauge mediated supersymmetry
breaking (GMSB) models (for a review see \citep{Giudice:1998bp})
there is a (gauge neutral) hidden sector which breaks supersymmetry
that is distinct from the moduli sector. Let us denote the fields
of this intermediate sector by indices $r,s,\ldots$. In our string
theory context they could be open string moduli. In addition such
models have a messenger sector which couples to this hidden sector
and is charged under the gauge group and since these are expected
to have only negligible F-terms we will denote them by the same indices
as the MSSM fields. The distinction between these sectors may be understood
in terms of the general formula \citep{Wess:1992cp} for the (unnormalized)
soft mass terms in the visible sector, \begin{equation}
\Delta M_{I\bar{J}}^{2}=-R_{I\bar{J}k\bar{l}}F^{k}F^{\bar{l}}-R_{I\bar{J}r\bar{s}}F^{r}F^{\bar{s}}-R_{I\bar{J}K\bar{L}}F^{K}F^{\bar{L}}+\frac{1}{3}F_{I}F_{\bar{J}}+K_{I\bar{J}}m_{3/2}^{2}.\label{eq:softmass}\end{equation}
 (Note for simplicity of exposition we have ignored mixed terms such
as $R_{I\bar{J}k\bar{L}}F^{k}F^{\bar{L}}$ above). The right hand
side of this equation is to be evaluated at the minimum of the scalar
potential. The general formula for the F-term is\begin{equation}
\bar{F}^{\bar{A}}=e^{K/2}K^{\bar{A}B}D_{B}W=e^{K/2}K^{\bar{A}B}(\partial_{B}W+K_{B}W).\label{eq:FA}\end{equation}
 The different mechanisms and mediations may be distinguished in terms
of the two physical scales $M_{P}\,(\simeq10^{18}GeV)\rightarrow1$
and the weak scale $G_{F}^{-1/2}\,(\simeq100GeV)\rightarrow10^{-16}.$
Now the F-terms of the visible sector fields $F^{I}$ are at most
of the order of the squared Higgs vacuum expectation value (vev) or
the Higgs vev times the gravitino mass i.e. $\sim10^{-30}$ or $10^{-15}m_{3/2}$
(whichever is larger). On the other hand tuning the cosmological constant
to zero implies\begin{equation}
F^{A}\bar{F}^{\bar{B}}K_{A\bar{B}}-3m_{3/2}^{2}=0.\label{eq:CCtune}\end{equation}
 This means that (given that the Kaehler metric is positive definite)
$|F^{A}|\lesssim m_{3/2}$. Now clearly the fourth term of (\ref{eq:softmass})
is much smaller (by a factor of $10^{-30}$) than $m_{3/2}^{2}$ and
so can be ignored (unless $m_{3/2}<10^{-30}\,(\sim10^{-6}eV)$ which
we assume is not the case.

The first term in (\ref{eq:softmass}) is the classical contribution
of the moduli which typically take Planck scale expectation values.
In string theory for instance, in order for the four dimensional low
energy approximation to be valid, these moduli must typically take
values which are somewhat larger than the Planck scale. The curvature
is of order one or less on the Planck scale so that the contribution
of this term to the squared soft mass is at most $O(m_{3/2}^{2})$.
This is called the moduli mediated (MMSB) contribution.

The second term can come from some hidden sector field (open string
modulus) which acquires an F-term as in GMSB. As argued earlier all
F-terms are $\lesssim O(m_{3/2})$. Classically the corresponding
moduli space curvature is at most order one (and typically in models
it is either zero or highly suppressed) and so this contribution would
not dominate over the classical contribution from the (closed string)
moduli sector. However there are wave function renormalization effects
which effectively enhance loop effects since the moduli space curvature
goes like $\phi^{-1}$ where $\phi$ is the lowest component of some
scalar field. Thus the contribution of this to the soft mass is effectively
like $|\epsilon F^{\phi}/\phi|^{2}$ (where $\epsilon=g^{2}/16\pi^{2}$
with $g$ a gauge coupling). This would be enhanced over the MMSB
contribution if the potential is such that $\phi$, the hidden sector
SUSY breaking field in GMSB, has a vev which is significantly smaller
than the Planck scale (which could be the case for open string moduli).
One does not really need sequestering in this case for the GMSB contribution
to dominate the modulus contribution - all that is needed is that
$F^{\phi}\le F\le(10^{-8})^{2}$ and $\phi\lesssim10^{-3}$. The former
inequality follows from (\ref{eq:CCtune}) and the fact that any classical
contribution to the soft masses from MMSB will be of order $m_{3/2}$
which should therefore be taken to be $O(10^{-16})\sim10^{2}GeV$,
while the latter is required so that the $O(\alpha/4\pi$) suppression
in the flavor conserving GMSB contribution is compensated. However
this gives a GMSB contribution at the same level as the MMSB contribution.
So unless MMSB already has suppressed FCNC (in which case there is
no need for any GMSB mechanism) we need to suppress the gravitino
mass well beyond the usual MMSB value. Thus in typical GMSB models
one has $F^{\phi}\sim(10^{9}GeV)^{2}$ or less and $\phi\sim10^{13}GeV$
or less with a gravitino mass (which in effect would be the size of
possibly FCNC violating MMSB contributions) of around a few GeV or
less.

The third mechanism is usually called anomaly mediated supersymmetry
breaking (AMSB) \citep{Randall:1998uk,Giudice:1998xp,Pomarol:1999ie}
and is supposedly associated with Weyl (or conformal) anomalies in
supergravity. As discussed in \citep{deAlwis:2008aq} (based on the
work of \citep{Dine:2007me}) this actually consists of two different
contributions. One of them arises from the Weyl anomaly of the theory.
This effect will be present even in the absence of matter fields -
for example in supergravity coupled to super-Yang-Mills fields. In
addition there is a contribution that arises from the mechanism pointed
out by Dine and Seiberg \citep{Dine:2007me} which in fact has nothing
to do with Weyl anomalies. This is like GMSB in that the contribution
to the soft masses arises from a quantum effect, but instead of having
an intermediate scale sector as in GMSB, it relies on the fact that
the Higgs field acquires a non-zero vev in the physical vacuum. This
in turn leads to an F-term for the Higgses of the form $F^{H}\sim m_{3/2}H$.
Now given that the classical contribution to the curvature is $O(1)$
or less, this gives a negligible contribution to the soft masses.
However there is a quantum contribution which gives a moduli space
curvature of the form $R\sim\epsilon^{2}/H^{2}$, giving a squared
soft mass of $O(\epsilon^{2}m_{3/2}^{2})$. But this is usually much
smaller than the contribution from MMSB and so the latter must be
suppressed, i.e. the classical moduli contribution to SUSY breaking
must be sequestered \citep{Randall:1998uk}, unless its FCNC effects
are negligible.

In this paper we will discuss a class of models which contain the
minimal inputs that are necessary to have soft supersymmetry breaking
terms in the MSSM, are consistent with the suppression of flavor violating
terms, and which can be embedded in a supergravity/string theoretic
framework. In section \ref{sec:The-Model} we discuss the hidden sector
which is responsible for supersymmetry breaking. This is the closed
string moduli sector of the theory. Obviously we cannot start with
a generic point on the landscape of string solutions since this will
not have the tiny cosmological constant that is observed. Also it
will most probably have large (i.e. string scale or Kaluza-Klein scale)
supersymmetry breaking, so that we would certainly not be led to the
MSSM, which in this bottom up approach is our starting point. Thus
we need to be restricted to those points in the landscape which have
a nearly zero cosmological constant and low energy supersymmetry breaking.
We focus on those models where this happens in the simplest possible
way. We will consider a moduli sector breaking supersymmetry in such
a way that it is not directly passed on to the visible sector at the
classical level. At one loop level there are quadratic divergences
(with a cutoff that we will identify with the GUT/KK scale). This
requires the retuning of the cosmological constant and it gives a
flavor diagonal contribution to the soft supersymmetry breaking parameters
that is proportional to $\frac{\Lambda^{2}}{16\pi^{2}}m_{3/2}^{2}$.
In general however there is a FCNC violating term which needs to be
suppressed by fine tuning the fluxes in an appropriate manner. In
effect this is a derivation of a quantum version of the mSUGRA model.
In addition to this there is the mechanism identified in \citep{Dine:2007me}\citep{deAlwis:2008aq}
which replaces what is usually presented as anomaly mediated supersymmetry
breaking (AMSB). Thus the basic claim of this paper is that the simplest
model of supersymmetry breaking that is consistent with all constraints
(both theoretical and phenomenological) and which is independent of
ad hoc uplift terms, is a version of mSUGRA which comes from an high
energy quantum effect, plus the low-energy quantum effect identified
in \citep{Dine:2007me}\citep{deAlwis:2008aq}.

\section{The Model\label{sec:The-Model}}

The moduli sector is taken to come from type IIB compactified on a
Calabi-Yau orientifold \citep{Giddings:2001yu} \citep{Kachru:2003aw}
with the visible sector being on a set of D3 branes. While a detailed
construction of such a model is not yet available it is very plausible
that one exists. Indeed it is likely that our arguments here apply
to a whole class of such models since only very generic properties
of such a construction are used. For simplicity we consider a model
with just one Kaehler modulus $T$ but a large number $h_{21}\gtrsim O(10^{2})$
of complex structure moduli $z^{\alpha}$, but it should be clear
from the discussion that an extension to compactifications with several
Kaehler moduli is straightforward. Also to stabilize the $T$ modulus
we will need non-perturbative terms as in KKLT \citep{Kachru:2003aw}.

The MSSM sector will have (schematically) quark/lepton $SU(2)$ doublet
superfields denoted by $Q^{i}/L^{i}$ and the corresponding singlet
conjugate fields $U^{ci}\, D^{ci},\, E^{ci}$ with $i$ being a family
index. The Higgs fields are $H_{u},\, H_{d}$. For the Kaehler potential
we take \begin{eqnarray}
K & = & -3\ln(T+\bar{T}-(H_{u}\bar{H_{u}}+H_{d}\bar{H_{d}}+z_{I\bar{J}}^{Q}Q^{I}\bar{Q}^{\bar{J}}+(x_{IJ}Q^{I}Q^{J}+h.c.))\nonumber \\
 &  & -\ln(S+\bar{S})-\ln k(z,\bar{z})\label{eq:Kfull}\\
 & = & K_{mod}+Z(T)_{I\bar{J}}\Phi^{I}\Phi^{\bar{J}}+\frac{1}{2}(X_{IJ}\Phi^{I}\Phi^{J}+h.c.)+\ldots\label{eq:Kfull1}\\
K_{mod} & = & -3\ln(T+\bar{T})-\ln(S+\bar{S})-\ln k(z,\bar{z}),\, Z_{I\bar{J}}=\frac{3z_{I\bar{J}}}{T+\bar{T}},\, X_{IJ}=\frac{3x_{IJ}}{T+\bar{T}}.\label{eq:KmodZ}\end{eqnarray}
 In the above $\Phi^{I},\, I=1,\ldots N_{v}$ denotes all the visible
sector fields. Note that in this model the space of dilaton-axion
$S$ and the complex structure moduli $\bar{z}^{\alpha}$, and the
space $T,\Phi^{I}$ are not direct product spaces and the metric is
not a direct sum of the metrics on these two spaces since $x_{IJ}$
is in general a function of the complex structure moduli, though it
vanishes when $z^{\alpha}=0$ \citep{Grana:2003ek}. Also the form
given in the first line of \eqref{eq:Kfull} is valid only to linear
order in the $z^{\alpha}$. While many authors (see for example \citep{DeWolfe:2002nn}\citep{Choi:2004sx},\citep{Kachru:2003sx})
use a direct sum form for the Kaehler potential in the presence of
D3 branes, this is only true if the complex structure moduli and the
dilaton are frozen at zero.

For the moduli superpotential we have\begin{equation}
W_{mod}=W_{flux}(S,z)+\sum_{n}A_{n}(S,z)e^{-a_{n}T},\label{eq:W0new}\end{equation}
 while for the MSSM superpotential we take \begin{equation}
W_{MSSM}=\tilde{\mu}H_{u}H_{d}+y_{uij}Q^{i}H_{u}U^{cj}+y_{Dij}Q^{i}H_{d}D^{cj}+y_{Eij}L^{i}H_{d}E^{cj}.\label{eq:Wmssm}\end{equation}
 In the above $S$ is the dilaton-axion superfield and $z=\{z^{\alpha}\},\,(\alpha=1,\ldots,h_{21})$
denotes the set of complex structure moduli and $T$ is the Kaehler
modulus of some Calabi-Yau orientifold (with $h_{11}=1$) compactification
of type IIB string theory. The first term in (\ref{eq:W0new}) comes
from internal magnetic fluxes and the second is a series of non-perturbative
(NP) terms coming from condensing gauge groups associated with D7-branes
\citep{Giddings:2001yu}\citep{Kachru:2003aw}. Also the MSSM sector
is located on a stack of D3 branes. For details of the dependence
of this superpotential on the closed string moduli see \citep{Camara:2003ku}
\citep{Grana:2003ek}. The model has a $R$-parity symmetry under
$\Phi(\theta)\rightarrow\pm\Phi(-\theta)$ with the plus sign for
the Higgses and minus sign for quark and lepton superfields. There
is also a PQ symmetry (if the $\mu$-term is set to zero) with charges
\begin{equation}
PQ:\, Q=L=U^{c}=D^{c}=L^{c}=-\frac{1}{2},\, H^{u}=H^{d}=1.\label{eq:PQ}\end{equation}
 and all moduli having zero charge. The moduli potential is\begin{equation}
V_{mod}=\frac{1}{k(z,\bar{z})(S+\bar{S})(T+\bar{T})^{2}}\{\frac{1}{3}(T+\bar{T})|\partial_{T}W_{mod}|^{2}-2\Re\partial_{T}W_{mod}\bar{W}_{mod}\}+|F^{S}|^{2}K_{S\bar{S}}+F^{z}F^{\bar{z}}k_{z\bar{z}}.\label{eq:Vmoduli}\end{equation}
 Now if one ignores quantum corrections, one would want to look for
a local minimum of this potential with zero cosmological constant
(CC) and SUSY breaking only in the $T$ direction, %
\footnote{One could of course look for more general minima where the SUSY breaking
is shared amongst all the moduli. However in practice in all models
discussed so far one usually looks for minima in which one starts
from the no-scale potential where $F_{T}\ne0,$ but the other F-terms
are zero and then expect the addition of the non-perturbative (NP)
terms that fix the T-modulus not to change this situation very much,
at least for large volume compactifications (see for example \citet{Balasubramanian:2005zx}).
Of course in the original toy model of KKLT $F_{T}=0$, but this was
a result of ignoring the non-trivial effects of the other moduli,
which really cannot be frozen in the presence of the NP terms. See
\citet{deAlwis:2005tf} for a discussion of these issues.%
} i.e. \begin{equation}
V_{mod}|_{0}=0,\, F|_{0}^{S}=F^{z}|_{0}=0,\, F^{T}|_{0}\ne0.\label{eq:susybreakmin}\end{equation}
There is certainly no obstruction to finding such a minimum and with
a sufficient number of complex structure moduli and non-perturbative
terms it is reasonable to expect that such a SUSY breaking minimum
exists. However the $T$ modulus - the scalar partner of the Goldstino
- has zero mass if the CC is fine-tuned exactly to zero. It should
be stressed though that this does \textit{not} imply that this modulus
is not stabilized, since we have included the non-perturbative terms
which are explicitly $T$ dependent. In other words the equation $\partial_{T}V=0$
will have a non-trivial solution because of the first term of \eqref{eq:Vmoduli}%
\footnote{The necessary conditions for stability in SUGRA models coming from
string theory were first discussed in \citep{Brustein:2004xn} and
were generalized by Gomez-Reino et al. \citep{Gomez-Reino:2006dk}\citep{Gomez-Reino:2006wv}\citep{Covi:2008ea}.%
}. 

In fact however as we will see in the next section the quantum corrections
would have required us to re-fine-tune the CC if we had started with
a zero value for it. So anticipating this what we really have to do
is to start at the classical level by fine-tuning the CC to be a small
(actually negative) value - much smaller, in absolute value, than
$m_{3/2}^{2}$. In this case there is certainly no obstruction to
having positive non-zero (squared) masses for all the moduli. Also
there will be additional contributions to the masses of the visible
sparticles, from the quantum corrections. In fact the sort of minimum
we will start with is like the one analyzed in the large volume scenario
of \citep{Balasubramanian:2005zx}%
\footnote{In the analysis of \citep{Balasubramanian:2005zx} an $\alpha'$ correction
is also included in the Kaehler potential though it is not really
essential for the demonstration of the existence of a minimum as such.
We can of course include this but have avoided doing so explicitly
for simplicity since it does not change the qualitative features that
we are discussing in this work.%
}. The only difference is that unlike in that case we have to fine-tune
$W_{flux}$ at the minimum (by adjusting the fluxes) to a very small
value, in order to get an intermediate volume scenario with the volume
of the internal manifold ${\cal V}\sim T^{3/2}\sim10^{3}$. We need
this to preserve a field theoretic description up to the unifcation
scale (which will be identified with the cut-off $\Lambda\sim10^{-2}$
in the quantum theory) while having a gravitino mass in the $10TeV$
range. Furthermore (as discussed in the next section) the quantum
contribution to the CC is $O(N\frac{\Lambda^{2}}{16\pi^{2}}m_{3/2}^{2})$
where $N$ is the number of chiral multiplets in the theory. Thus
we expect a broken supersymmetric minimum $F|_{0}^{S}=F^{z}|_{0}=0,\, F^{T}|_{0}\ne0$,
with a small negative cosmological constant $-|V_{0}|$ such that

\begin{equation}
|V_{0}|\sim\frac{m_{3/2}^{2}}{{\cal V}}\sim O(N\frac{\Lambda^{2}}{16\pi^{2}}m_{3/2}^{2})\ll O(m_{3/2}^{2}).\label{eq:finetune}\end{equation}
It should also be stressed here that our framework does not need any
ad hoc uplift terms to get an acceptable value for the CC. This will
come about as a result of the fine tuning of the classical SUGRA CC
against the quantum effects that are discussed below.

The curvature component relevant to the soft mass calculation in this
model is $R_{T\bar{T}I\bar{J}}=\frac{1}{3}K_{T\bar{T}}Z_{I\bar{J}}+O(H^{2})$
so that using the standard expression for soft masses, given for example
in \citep{Kaplunovsky:1993rd}\citep{Brignole:1997dp}, we have\begin{equation}
m_{I\bar{J}}^{2}=m_{3/2}^{2}Z_{I\bar{J}}-F^{T}F^{\bar{T}}R_{T\bar{T}I\bar{J}}\sim O(m_{3/2}^{2}\frac{\Lambda^{2}}{16\pi^{2}})\ll m_{3/2}^{2}.\label{eq:m2<<}\end{equation}
 Similarly both the $B\mu$ and the trilinear couplings - the $A$-terms
- are also suppressed %
\footnote{If we had fine-tuned the CC exactly to zero at the classical level
we would have got zero for the soft masses and the $\mu$, $B\mu$
and $A$ terms as in the no-scale model.%
}. In the next two sections we will consider the quantum effects.

\section{Quadratic divergence issues and mSUGRA\label{sec:Quadratic-divergence-issues}}

It is well known that quadratic divergences are absent in (spontaneously
broken) global supersymmetry, but this is not really relevant for
phenomenology for well-known reasons. Any mechanism of supersymmetry
breaking (such as say dynamical SUSY breaking) is incomplete unless
it is embedded within supergravity. Then one needs to confront the
problem of quadratic divergences. In the following we will discuss
how the cosmological constant and the soft supersymmetry breaking
parameters get affected by these divergences.

\subsection{The Cosmological Constant and the Soft Masses}

To one-loop order but keeping only the $O(\Lambda^{2})$ (where $\Lambda$
is the cutoff) corrections we have the following \citep{Gaillard:1993es}\citep{Ferrara:1994kg}\citep{Choi:1997de})
formulae for the potential (at a minimum) and the (unnormalized) soft
mass terms.

\begin{eqnarray}
V|_{0} & = & (F^{m}\bar{F}^{\bar{n}}K_{m\bar{n}}-3m_{3/2}^{2})(1+\frac{(N-5)\Lambda^{2}}{16\pi^{2}})+\frac{\Lambda^{2}}{16\pi^{2}}(m_{3/2}^{2}(N-1)-F^{T}\bar{F}^{\bar{T}}R_{T\bar{T}}),\label{eq:V1loop}\\
m_{I\bar{J}}^{2} & = & V|_{0}Z_{I\bar{J}}+(m_{3/2}^{2}Z_{I\bar{J}}-F^{T}F^{\bar{T}}R_{T\bar{T}I\bar{J}})(1+\frac{(N-5)\Lambda^{2}}{16\pi^{2}})\nonumber \\
 &  & -\frac{\Lambda^{2}}{16\pi^{2}}[m_{3/2}^{2}R_{I\bar{J}}+m_{3/2}(F^{T}D_{T}R_{I\bar{J}}+F^{\bar{T}}D_{\bar{T}}R_{I\bar{J}})+F^{T}F^{\bar{T}}(D_{T}D_{\bar{T}}R_{I\bar{J}}\nonumber \\
 &  & -R_{\bar{T}}^{\,\,\bar{T}}R_{T\bar{T}I\bar{J}}-R_{T}^{\,\, T}R_{T\bar{T}I\bar{J}}+R_{I}^{\,\, K}R_{T\bar{T}K\bar{J}})].\label{eq:m2oneloop}\end{eqnarray}
 Here $N$ is the total number of chiral scalar superfields. In writing
these expressions we have kept, in the one loop correction terms,
the classical fine tuning values (\ref{eq:susybreakmin}) of the F-terms.

In estimating these corrections we will take the cutoff to be \begin{equation}
\Lambda\sim M_{GUT}\sim M_{KK}\sim10^{16}GeV\sim10^{-2}M_{P}\rightarrow\frac{\Lambda^{2}}{16\pi^{2}}\sim10^{-6}M_{P}^{2}.\label{eq:cutoff}\end{equation}
 The first question that needs to be addressed is the fine-tuning
of the cosmological constant. With the classical fine tuning (\ref{eq:susybreakmin})
and using $R_{T\bar{T}}\simeq\frac{1}{3}(N_{v}+2)K_{T\bar{T}}$ (where
$N_{v}$ is the number of visible sector fields) we would obtain at
one-loop a CC of order $\frac{\Lambda^{2}}{16\pi^{2}}m_{3/2}^{2}(N-N_{v}-3)=10^{-6}m_{3/2}^{2}M_{P}^{2}(h_{21}-1)$.
Since we need the number of complex structure moduli to be of $O(10^{2})$
in order to be able to fine tune the classical CC, this one loop correction
leads to a CC (assuming that the gravitino is at least of order the
SUSY mass splittings $10^{2-3}GeV$) that is a factor $\sim10^{86}$
too large! Thus as we discussed before we need to change the classical
starting point which ignored the fact that there are quantum corrections
\footnote{For a discussion of the consequences for string phenomenology of this
refinetuning problem see \citep{deAlwis:2006nm}.%
}. In other words to cancel the CC to the leading order in $N\Lambda^{2}/16\pi^{2}$
we need to add corrections to (\ref{eq:susybreakmin}) and (\ref{eq:finetune})
such that (with $M_{P}=1$) \begin{equation}
3m_{3/2}^{2}-F^{m}F^{\bar{n}}K_{m\bar{n}}=\frac{\Lambda^{2}}{16\pi^{2}}(m_{3/2}^{2}(N-1-(2+N_{v}))=\frac{\Lambda^{2}}{16\pi^{2}}m_{3/2}^{2}(h_{21}-1).\label{eq:refinetune}\end{equation}
 Note that since the RHS of this equation is positive the classical
CC (the negative of the LHS) would have to be negative. In this case
there is no obstruction to getting a positive squared mass for the
$T$ modulus and generically it will be $O(m_{3/2}^{2})$.

The actual minimum around which we work in calculating the soft masses
will also change the values of the F-terms of the moduli from those
given in \eqref{eq:susybreakmin} to the following (with $|F^{i}|\equiv\sqrt{K_{i\bar{i}}F^{i}F^{\bar{i}}}$):\begin{equation}
|F^{T}|=\sqrt{3}m_{3/2}+O(h_{21}\frac{\Lambda^{2}}{16\pi^{2}}m_{3/2}),\,|F^{S}|\lesssim O(\frac{\Lambda}{4\pi}m_{3/2}),\,|F^{z}|\lesssim O(\frac{1}{\sqrt{h_{21}}}\frac{\Lambda}{4\pi}m_{3/2}).\label{eq:Fnew}\end{equation}
We will thus assume that one can find such a minimum by adjusting
fluxes and there certainly is no obstruction to doing so. 

Now let us calculate the soft masses by including the quantum corrections.
The first term in (\ref{eq:m2oneloop}) has now been re-fine-tuned
to zero. However the second term is no longer zero and there is an
additional contribution from the third term. To calculate these we
need the curvatures derived from the Kaehler potential (\ref{eq:Kfull}):
\begin{eqnarray*}
R_{T\bar{T}I\bar{J}} & = & \frac{K_{T\bar{T}}z_{I\bar{J}}}{T+\bar{T}}+O(\Phi^{2}),\, R_{I\bar{J}K\bar{L}}=\frac{3}{(T+\bar{T})^{2}}(z_{I\bar{J}}z_{K\bar{L}}+z_{I\bar{L}}z_{K\bar{J}}-z_{IK}z_{\bar{L}\bar{K}})+O(\Phi^{2})\\
R_{I\bar{J}} & = & \frac{N_{v}+1}{T+\bar{T}}z_{I\bar{J}}+O(\Phi),\, D_{T}R_{I\bar{J}}=O(\Phi),\, D_{T}D_{\bar{T}}R_{I\bar{J}}=O(\Phi)\\
R_{I}^{\,\, K}R_{T\bar{T}K\bar{J}} & = & \frac{N_{v}+1}{(T+\bar{T})^{3}}z_{I\bar{J}}+O(\Phi^{2}).\end{eqnarray*}
 So (given that the MSSM fields $\Phi$ have values that are highly
suppressed $<O(10^{-16})$ relative to the Planck scale) we find from
(\ref{eq:m2oneloop}) on using (\ref{eq:refinetune}) that the the
largest quantum contribution to the soft mass squared is\begin{equation}
m_{I\bar{J}}^{'2}\sim(h_{21}-2N_{v})\frac{\Lambda^{2}}{16\pi^{2}}m_{3/2}^{2}Z_{I\bar{J}}\sim(h_{21}-2N_{v})10^{-6}m_{3/2}^{2}Z_{I\bar{J}},\label{eq:m2correction}\end{equation}
 and is positive provided that $h_{21}>2N_{v}$. It is also flavor
diagonal. In fact it is precisely of the form assumed by mSUGRA models
of supersymmetry breaking and is easily obtained for generic Calabi-Yau
orientifold compactifications.

The flavor conserving two loop quantum corrections coming from fluctuations
of light fields that we will consider in the next section, are in
fact of the same order provided that the number of complex structure
moduli is $O(10^{2})$. In fact since $N_{v}$ is also of the same
order, this is a necessary condition to get positive squared masses.
Of course the tuning of the cosmological constant already requires
the number of cycles in the compactification manifold to be at least
of this order. Thus this contribution is $O(10^{-4}m_{3/2}^{2})$.
However if this (\ref{eq:m2correction}) had been flavor violating
the model (even with the flavor conserving effect of the next section)
would have been in danger of being ruled out since the flavor violating
effects need to be down by a factor of around $10^{-3}$ compared
to the flavor conserving one. Note that the flavor conservation of
the soft masses calculated in this section is entirely due to the
fact that the visible sector field space metric factorizes into a
modulus dependent factor and a matrix in generation space. This in
turn is a reflection of the fact that all visible fields are from
a stack of D3 branes. This would not have been the case if the visible
sector came partially from D3 branes and partially from (wrapped)
D7 branes for instance. Such a general embedding would have resulted
in a metric $Z_{I\bar{J}}=f(M,\bar{M})z_{I\bar{J}}+g(M,\bar{M})z'_{I\bar{J}}$
where $M$ denotes the set of moduli and the dilaton and $z_{I\bar{J}},z'_{I\bar{J}}$
are in general different matrices so that the curvature would not
have been proportional to $Z_{I\bar{J}}$ , and hence we would have
had flavor changing terms at an unacceptable level.

Nevertheless it should be noted that the above calculation of curvatures
are done in the linearized (in the complex structure moduli $z^{\alpha}$
) solution given in \citep{Grana:2003ek}. It is indeed possible that
the complete solution will yield a contribution to \eqref{eq:m2correction}
that is not proportional to the matrix $Z_{I\bar{J}}$ and so in general
will lead to fine tuning. For instance in general $z_{I\bar{J}}$
would be a function of the complex structure moduli $z^{\alpha},\bar{z}{}^{\beta}$
and so there would be a contribution to the Ricci tensor in the MSSM
directions of the form $R_{I\bar{J}}\sim K^{\alpha,\bar{\beta}}\partial_{\alpha}\partial_{\bar{\beta}}Z_{I\bar{J}}$
which is not (in general) proportional to $ $$Z_{I\bar{J}}$. This
could in principle give, from the second line of \eqref{eq:m2oneloop}
a contribution as large as the one in \eqref{eq:m2correction}. In
this case we need additional fine-tuning to 1 part in $10^{3}$ to
achieve the necessary suppression of FCNC and this can be done by
appropriate choices of the fluxes which determine the complex structure
moduli.

It should be noted that (given the suppression of classical soft terms
in our model) (\ref{eq:m2correction}) by itself would give soft mass
terms at an acceptable level provided that the gravitino mass is a
factor of $10^{2}$ larger than the soft mass - i.e. we would need
a gravitino with $m_{3/2}\gtrsim10TeV$. This is typical of so-called
AMSB scenarios where the classical terms are {}``sequestered'' \citep{Randall:1998uk}
as is the case with our classical starting point (\ref{eq:Kfull}).
The point of our discussion here is to show that the quadratic divergences
that are inevitably present, give a contribution which is competitive
with the `AMSB' effects. 

As for the $A$ terms, adding the quadratically divergent one-loop
effects gives \begin{equation}
A_{IJK}=e^{K_{m}/2}\frac{W_{m}^{*}}{|W_{m}|}\{F^{i}D_{i}y_{IJK}(1+\frac{N-5}{16\pi^{2}}\Lambda^{2})-\frac{\Lambda^{2}}{16\pi^{2}}O(F^{T})\}\label{eq:Aquad}\end{equation}
 where the sum in the first term in parentheses excludes the $T$
modulus (recall that the classical contribution is suppressed since
in the no-scale model it would be exactly zero while here it is $O(\Lambda^{2}/16\pi^{2})$).
The second term consists of terms that are proportional to $y_{IJK}$.
As shown in \citep{Grana:2003ek} the first term is proportional to
$y_{IJK}$ and hence when (due to quantum effects in our case) the
$F^{i}$ are turned on, no significant flavor violating effects are
generated.

\subsection{Consistency Issues}

Let us now check what the cancellation of the one-loop contribution
to the cosmological constant implies for the F-terms of the moduli.
Using \eqref{eq:W0new} we have (assuming for simplicity that there
is only one NP term)\begin{equation}
F^{\bar{T}}=e^{K/2}K^{\bar{T}T}D_{T}W=e^{K/2}K^{\bar{T}T}(-aAe^{-aT}+K_{T}W)\label{eq:FT}\end{equation}
Note that in this formula as well as in the arguments in the rest
of this subsection the values of the moduli are understood to be taken
at the local (negative CC) minimum of section II.

The requirement that the one loop CC contribution to the CC is cancelled
then yields \begin{equation}
3m_{3/2}^{2}-K_{T\bar{T}}F^{T}F^{\bar{T}}=\frac{2\sqrt{3}a\Re Ae^{-aT}}{k^{1/2}(S+\bar{S})^{1/2}(T+\bar{T})^{1/2}}m_{3/2}+O(e^{-2aT})\sim h_{21}\frac{\Lambda^{2}}{16\pi^{2}}m_{3/2}^{2},\label{eq:consis1}\end{equation}
 where in the last relation we have used \eqref{eq:refinetune}. This
gives us an estimate of how large the non-perturbative contribution
(at the minimum) should be:\begin{equation}
Ae^{-aT}\sim a^{-1}(T+\bar{T})^{1/2}h_{21}\frac{\Lambda^{2}}{16\pi^{2}}m_{3/2}.\label{eq:NPterm}\end{equation}
 Let us check now that this gives a reasonable value for $a$. First
we need to estimate the value of $T$. Using the fact that the Kaluza-Klein
mass $M_{KK}$ in Planck units is $1/T$ %
\footnote{For a discussion of the scales involved in both the unwarped and warped
cases see \citep{deAlwis:2006am}. Note that we are actually discussing
a class of type IIB solutions where warping can be ignored. It is
completely unclear to us how to use the SUGRA formalism when warping
is significant.%
}, we have\begin{equation}
\frac{1}{T}\sim M_{KK}\sim\Lambda\sim10^{-2}\Longrightarrow T\lesssim O(10^{2})\label{eq:Testimate}\end{equation}
 Assuming $A\sim O(1)$ and $m_{3/2}\sim10TeV\sim10^{-14}M_{P}$ from
\eqref{eq:NPterm} we estimate $a\gtrsim O(1/10)$ which is a reasonable
value since it would correspond to condensing gauge groups %
\footnote{In the KKLT picture this would come from the gauge theory on seven-branes
wrapping a 4-cycle in the internal manifold.%
} of rank $N\sim10-100$.

Let us ask how big the F-component of the complex structure moduli
and the dilaton can be. In the presence of both imaginary anti-self-dual
(IASD) fluxes (in the terminology of GKP \citep{Giddings:2001yu})
and non-perturbative terms we have \begin{eqnarray}
F^{\bar{\alpha}} & = & K^{\bar{\alpha}\beta}e^{K/2}(D_{\beta}W_{flux}+K_{\beta}Ae^{-aT})=K^{\bar{\alpha}\beta}e^{K/2}(I_{\beta}+K_{\beta}Ae^{-aT})\label{eq:zIASD}\\
F^{S} & = & K^{\bar{S}S}e^{K/2}(D_{S}W_{flux}+K_{S}Ae^{-aT})=K^{\bar{S}S}e^{K/2}(I+K_{S}Ae^{-aT})\label{eq:SIASD}\end{eqnarray}
 Here $I_{\beta}$ is an (2,1) flux and $I$ is a (3.0) flux. Now
the classical solution (in the absence of NP terms) requires that
these IASD fluxes are zero. In finding a minimum for the classical
potential that includes the non-perturbative terms such that the one-loop
CC is cancelled, it is clear that we should not turn on IASD fluxes,
since these would generically give large positive terms in the potential
and violate the last two relations in \eqref{eq:Fnew}. In that case
using the estimate \eqref{eq:NPterm} we have\begin{eqnarray}
F^{\bar{\alpha}} & \sim & \frac{K^{\bar{\alpha}\beta}K_{\beta}}{k^{1/2}(S+\bar{S})2a\Re T}h_{21}\frac{\Lambda^{2}}{16\pi^{2}}m_{3/2},\label{eq:Falpha}\\
F^{\bar{S}} & \sim & \frac{K^{\bar{S}S}K_{S}}{k^{1/2}(S+\bar{S})2a\Re T}h_{21}\frac{\Lambda^{2}}{16\pi^{2}}m_{3/2}.\label{eq:FS}\end{eqnarray}
 Finally we observe that for consistency these values of the F-terms
of these moduli implies that their masses are considerably lower than
the string scale. This can be seen by imposing the constraint that
the mass of the scalar partner of the Goldstino should be of the order
of $m_{3/2}$. Defining the unit vector in the Goldstino direction
$u^{m}\equiv F^{m}/\sqrt{K_{m\bar{n}}F^{m}F^{\bar{n}}}$ we have

\begin{eqnarray*}
u^{T} & = & \frac{F^{T}}{\sqrt{F^{T}F_{T}(1+\epsilon^{2})}}\sim\frac{e^{i\phi_{T}}}{\sqrt{K_{T\bar{T}}(1+\epsilon^{2})}}\\
u^{\alpha} & = & \frac{F^{\alpha}}{\sqrt{F^{T}F_{T}(1+\epsilon^{2})}}\sim\frac{\epsilon^{\alpha}e^{i\phi_{\alpha}}}{\sqrt{K_{T\bar{T}}(1+\epsilon^{2})}}\end{eqnarray*}
 where $\epsilon^{\alpha}\equiv|F^{\alpha}|/\sqrt{F^{T}F_{T}}$ and
$\epsilon^{2}=\epsilon^{\alpha}\epsilon_{\alpha}$ and we take $\alpha=0,1,\ldots h_{12}$
with the index $\alpha=0$ identified with the dilaton $S$. So for
the squared mass of the sGoldstino we have \[
u^{m}V_{m\bar{n}}u^{\bar{n}}=\frac{K^{T\bar{T}}V_{T\bar{T}}}{1+\epsilon^{2}}+\frac{\epsilon^{\alpha}V_{\alpha\bar{\beta}}\epsilon^{\bar{\beta}}}{1+\epsilon^{2}}\sim O(m_{3/2}^{2})\]
 This tells us that $\epsilon^{\alpha}\sim m_{3/2}/m_{\alpha}$ so
that we have the result%
\footnote{This has also been obtained in \citep{Choi:2008hn} though the argument
there is different from the above and appears to depend on global
SUSY.%
}cd \[
F^{\alpha}\sim\frac{m_{3/2}}{m_{\alpha}}m_{3/2}.\]
 Comparing with \eqref{eq:Falpha}\eqref{eq:FS} we see that this
implies $m_{S}\sim m_{z}\sim10^{4}m_{3/2}$.

\subsection{$\mu$ and $B\mu$ terms}

The expression for the effective $\mu$ term (after integrating out
the moduli) is given by (see for example \citep{Kaplunovsky:1993rd,Choi:1997de}
and references therein) \begin{equation}
\mu_{IJ}=e^{K_{mod}/2}\tilde{\mu}_{IJ}+m_{3/2}X_{IJ}-\bar{F}^{\bar{A}}\partial_{\bar{A}}X_{IJ}+O(\frac{\Lambda^{2}}{16\pi^{2}}m_{3/2}).\label{eq:mueff}\end{equation}
 In this expression the second and third term are of the order of
the supersymmetry breaking but there is no reason for first term (which
comes from the original superpotential) to be of the same order -
generically it would be $O(1)$ in Planck units. That of course would
be a disaster since in that case there would be no electroweak symmetry
breaking. This is the well known $\mu$ problem of the MSSM.

In our string theory based model of gravity mediated SUSY breaking
with the MSSM located on D3 branes however $\tilde{\mu}=0$ $ $and
the effective $\mu$ term emerges from the well-known Giudice-Masiero
effect \citep{Giudice:1988yz}. As shown by Grana et al \citep{Grana:2003ek}
$\mu_{IJ}=-\bar{F}^{\bar{\alpha}}\partial_{\bar{\alpha}}X_{IJ}$ so
using \eqref{eq:Falpha} and the fact that the sum over $\alpha$
has $h_{21}$ terms, we get \begin{equation}
\mu\sim O(\frac{h_{21}^{2}}{aT}\frac{\Lambda^{2}}{16\pi^{2}}m_{3/2})\sim10^{-2}m_{3/2},\label{eq:mugrana}\end{equation}
 where we have used the value $aT\sim O(10)$ (see line after \eqref{eq:Testimate})
and $h_{21}\sim3\times10^{2}$.

Also using the calculation of the $B\mu$ term given in \citep{Grana:2003ek}
we have \begin{equation}
B\mu_{IJ}=V_{classical}|_{0}X_{IJ}\sim O(h_{21}\frac{\Lambda^{2}}{16\pi^{2}}m_{3/2}^{2})\sim10^{-3}m_{3/2}^{2},\label{eq:Bmugrana}\end{equation}
 so that \begin{equation}
\frac{B\mu}{\mu}\sim\frac{aT}{h_{21}}m_{3/2}\sim3\times10^{-2}m_{3/2}.\label{eq:Bmubymu}\end{equation}

\subsection{Gaugino mass}

Let us now consider the gaugino masses. The general formula for these
is

\begin{equation}
m_{a}=\frac{1}{2}(\Re f_{a})^{-1}F^{A}\partial_{A}f_{a}(\Phi),\label{eq:gauginomass}\end{equation}
 and we will only get a contribution if the gauge coupling function
$f_{a}$ depends on a chiral multiplet that acquires a non-vanishing
F-term. In our case since the gauge theory on the D3 branes is independent
of the moduli of the internal manifold and so the gaugino mass is
suppressed relative to the gravitino mass. In particular the quadratic
divergence in the potential led us to shift the minimum resulting
in the new F-term values (\ref{eq:refinetune}). Since the gauge coupling
function on D3 branes depends (in the Einstein frame) on the dilaton
this gives a non-vanishing contribution (since $f\sim S\sim1/g^{2}$)
\begin{equation}
\frac{m_{a}}{g_{a}^{2}}=\frac{1}{2}F^{S}\partial_{S}f_{a}(S)\sim O\left(h_{21}\frac{\Lambda^{2}}{16\pi^{2}}m_{3/2}\right)\lesssim10^{-3}m_{3/2}.\label{eq:ma1}\end{equation}

\subsection{mSUGRA parameters}

As discussed above we need to choose the cutoff $\Lambda\sim10^{-2}$
and we took $h_{21}\sim3\times10^{2}$. Taking $m_{3/2}\sim10^{4}GeV$
we get reasonable soft parameters except that the gaugino masses are
too small. But as we shall see in the next section there is an `AMSB'
contribution to the gaugino masses which is much larger than the mSUGRA
contribution. Our mSUGRA parameters are,\begin{equation}
\mu\sim10^{-2}m_{3/2}\sim100GeV,\, m_{s}\sim2\times10^{-2}m_{3/2}\sim200GeV,\,\frac{B\mu}{\mu}\sim\frac{aT}{h_{21}}m_{3/2}\sim300GeV.\label{eq:msugraparam}\end{equation}
 Note that a somewhat larger value of the gravitino mass (say $\sim30TeV$
as in typical `AMSB' scenarios ) would also give acceptable values
provided we take $h_{21}$ to be a little larger for example $h_{21}\sim4-5\times10^{2}$.
The gaugino masses however would still be of $<O(100GeV)$ and hence
if this is the only contribution the model would be ruled out on phenomenological
grounds. However as we discuss in the next section there are additional
contributions.

\section{SUSY breaking and AMSB\label{sec:DS-susy-breaking}}

In the previous section we showed how mSUGRA like SUSY breaking terms
arise at the cutoff scale $\Lambda$, in a model which can be naturally
embedded in a type IIB string theoretic setup. These boundary conditions
of mSUGRA need to be evolved down to the electro-weak scale in order
to evaluate the actual predictions of this set up for the MSSM parameters.
This calculation is just the same as in the usual mSUGRA set up and
we will not go over it.

However there is a new contribution to any such theory that needs
to be considered. This is usually assumed to be due to conformal anomalies
and is referred to as anomaly mediated supersymmetry breaking (AMSB)
\citep{Randall:1998uk}\citep{Giudice:1998xp,Pomarol:1999ie}\citep{Bagger:1999rd}.
The most detailed SUGRA based derivation of the gaugino mass is given
in the last citation and (in our conventions) reads\begin{equation}
\frac{m_{a}}{g_{a}^{2}}=\Re[F^{i}\partial_{i}f_{a}(\Phi)|-\frac{1}{8\pi^{2}}(b'_{a}m_{3/2}+c_{a}F^{i}\partial_{i}K_{m}+2T_{R}F^{i}\partial_{i}\ln Z_{r})],\label{eq:Bagger}\end{equation}
where $c_{a}=T(G_{a})-\sum_{r}T_{a}(r)$ and $b'_{a}=3T(G_{a})-\sum_{r}T_{a}(r)$
with $T(G_{a}),T_{a}(r$), being the traces of a squared gauge group
generator in the adjoint and a matter representation $r$ respectively.
The sum over representations go over all states which are effectively
massless at the cut off scale. In our approximately no-scale model
with the MSSM on a stack of D3 branes the contribution of the first
term was given in \eqref{eq:ma1}. We also see that there is a cancellation
amongst the terms in the paranthesis in the second term in LHS of
\eqref{eq:Bagger} so that we effectively get from this formula the
same result as before, namely\begin{equation}
\frac{m_{a}}{g_{a}^{2}}\sim O(10^{-3}m_{3/2}).\label{eq:Bagger2}\end{equation}
If correct this would mean that the gaugino masses are well below
the experimental upper limit, even for gravitino masses that are as
high as $100TeV$, which is the highest one can tolerate without seriously
affecting the hierarchy. This would imply that type IIB string theory
with the MSSM on D3 branes can only give a split supersymmetry type
of scenario. Thus we would need $m_{3/2}\sim10^{3}TeV$, giving gaugino
masses $m_{a}\sim1TeV$, but soft masses as well as $\mu$ and $B\mu/\mu$
would then be $O(10TeV$) and the Higgs squared mass would be fine
tuned (at a level of 1 part in $10^{4}$). However as we will argue
below this conclusion is not warranted.

The point is that as shown in \citep{deAlwis:2008aq} the arguments
in \citep{Randall:1998uk}\citep{Giudice:1998xp,Pomarol:1999ie}\citep{Bagger:1999rd}
need to be revised. Let us briefly summarize this discussion. The
most important point is that the the so-called Weyl (or conformal)
compensator chiral superfield $C$ is a (non-propagating) field, and
the theory needs to have enough gauge freedom so that it can for instance
be set equal to unity to get the standard formulation of SUGRA. The
Weyl anomaly at one-loop effectively prevents this, and Kaplunovsky
and Louis (KL) \citep{Kaplunovsky:1994fg} showed by a careful and
detailed calculation how this anomaly could be cancelled thereby restoring
the gauge symmetry. Their discussion led to a corrected form for the
gauge coupling function in superspace (at the cutoff scale $\Lambda$) 

\begin{equation}
H_{a}(\Phi;\Lambda)=f_{a}(\Phi)-\frac{3b'_{a}}{4\pi^{2}}\ln C-\frac{T_{a}(r)}{4\pi^{2}}\ln(e^{-\frac{1}{3}K_{m}}Z_{r})|_{holomorphic}.\label{eq:H}\end{equation}
Projecting the F-term of this gives us the formula\begin{equation}
\frac{m_{a}}{g_{a}^{2}}=\Re[F^{i}\partial_{i}f_{a}(\Phi)|-\frac{b'_{a}}{8\pi^{2}}\frac{F^{C}}{C}-\frac{T_{a}(r)}{4\pi^{2}}F^{i}\partial_{i}(\ln(e^{-\frac{1}{3}K_{m}}Z_{r})],\label{eq:KL}\end{equation}
where $C,F^{C}$ are the lowest and highest components of the Weyl
compensator superfield. The question is what is the value of the second
term on the RHS. As shown by KL, in the Kaehler-Einstein frame (which
is the correct `physical' frame in which standard SUGRA low energy
results should be derived) $\frac{F^{C}}{C}=\frac{1}{3}K_{i}F^{i}$.
Putting this in (\ref{eq:KL}) we have\begin{eqnarray}
\frac{m_{a}}{g_{a}^{2}} & =\Re H_{a}(\Phi;\Lambda)|_{F} & =\Re[F^{i}\partial_{i}f_{a}(\Phi)|-\frac{c_{a}}{8\pi^{2}}F^{i}\partial_{i}K_{m}-\frac{T_{a}(r)}{4\pi^{2}}F^{i}\partial_{i}(\ln Z_{r})],\nonumber \\
 &  & =O(10^{-3}m_{3/2})-\frac{b_{a}'}{8\pi^{2}}m_{3/2}+O(\frac{\Lambda}{4\pi}m_{3/2}).\label{eq:KL2}\end{eqnarray}
In the last line the first term is from \eqref{eq:ma1}, and we have
used the values for the Kahler potential and F-terms for our model
from \eqref{eq:KmodZ}\eqref{eq:W0new}\eqref{eq:Falpha}\eqref{eq:FS}.
This is the correct contribution from AMSB and indeed it gives a value
for the gaugino masses (with $m_{3/2}\sim30TeV)$ that is of the right
order.

However as shown in \citep{Dine:2007me} (DS) and elaborated on in
\citep{deAlwis:2008aq} there is an additional contribution which
has nothing to do with Weyl anomalies but is a quantum effect in the
effective action. This arises as follows. The gauge coupling function
at the high (GUT) scale $\Lambda$ is given by the superfield \eqref{eq:H}.
At a low scale $\mu$ the one-loop beta function formula gives\begin{equation}
H_{a}(\Phi;\mu)=H_{a}(\Phi;\Lambda)-\frac{b'_{a}}{8\pi^{2}}\ln\frac{\Lambda}{\mu}.\label{eq:HRG}\end{equation}
Now assume that there is an intermediate threshold which is generated
by a superfield $X$ (and its F-term) acquiring non-zero values in
the ground state of the theory. The appropriate replacement of \eqref{eq:HRG}
in the Wilsonian effective action at the low scale $\mu$ is \begin{equation}
H_{a}(\Phi,X;\mu)=H_{a}(\Phi;\Lambda)-\frac{b_{a}}{8\pi^{2}}\ln\frac{X}{\mu}-\frac{b'_{a}}{8\pi^{2}}\ln\frac{\Lambda}{X}\label{eq:HRGX}\end{equation}
where $b_{a}$ is the beta-function coefficient below the scale set
by the vev $X_{0}$ of $X$. It may be obtained by integrating the
one-loop beta function above and below the scale set by $X$ and then
using holomorphy. In effect this is the usual argument that the superspace
gauge coupling function is one-loop exact. This is basically the argument
given in DS\citep{Dine:2007me} except that there the scale $\mu$
was not introduced and $X_{0}$ was eventually taken to zero. However
that would clearly introduce infra-red divergences and in any case
we should be in the Higgs phase where $X$ has a non-zero expectation
value and as DS argued their discussion really applies only in the
Higgs phase. It should also be noted that this formula is exact for
the Wilsonian coupling function. A similar formula is given in \citep{Giudice:1997ni}
in the context of gauge mediated supersymmetry breaking and is used
in \citep{Pomarol:1999ie} to argue for what is often called deflected
anomaly mediation in the literature%
\footnote{Formulae for gaugino masses, given in the literature on GMSB, which
have corrections which effectively involve $D^{2}X$ etc (where $D$
is the supercovariant spinor derivative) would necessarily take us
to a higher derivative theory and in any case are not relevant for
the Wilsonian coupling function that we are concerned with here. In
other words to the extent that we confine ourselves to a two derivative
Wilsonian action, the formula (\ref{eq:HRGX}) is exact.%
}. Indeed as pointed out in both \citep{Giudice:1997ni}\citep{Dine:2007me}
the $X$ dependence in this formula follows from holomorphy and the
necessity of reproducing the correct chiral anomaly from states that
have been integrated out to get the effective theory at scales below
that set by $X$. 

Taking the F-term of this and replacing the field and its F-term by
their ground state values, we have for the gaugino mass at the scale
$\mu\rightarrow X_{0}$\begin{equation}
\frac{m_{a}}{g_{a}^{2}}=\Re H_{a}(\Phi;\Lambda)|_{F}-\frac{b_{a}-b'_{a}}{8\pi^{2}}\frac{F^{X}}{X_{0}},\label{eq:gauginomu1}\end{equation}
\footnote{We note that the calculation is similar to that given in \citep{Giudice:1997ni}
the main difference being that the effective messenger scale is the
TeV scale and there is no messenger sector. It should also be noted
that the GMSB constraint $F^{X}/X_{0}^{2}<1$ which applies from the
necessity of ensuring non-negative messenger squared masses does not
apply here. First of all there are no messengers but more importantly
the mass formulae are given by the supergravity expression (\ref{eq:softmass})
and not by global SUSY formulae which are used in the GMSB context
where the gravitino mass contribution maybe ignored.%
}. In our model the only possible threshold below the GUT scale is
the weak scale set by the Higgs field itself. Thus we should take
the gauge neutral field $X$ to be the gauge neutral combination of
the two MSSM Higgs superfields, i.e. we may put $X^{2}=H_{u}H_{d}$.
In effect this was what was done by Dine and Seiberg in their toy
model outlining this general idea in \citep{Dine:2007me} and the
above mentioned chiral anomaly is in the global symmetry under which
both $H_{u}$ and $H_{d}$ rotate by the same phase and the quark
and lepton fields rotate by half the opposite phase - a symmetry which
is explicitly broken by the $\mu$ term. $X$ will acquire a non-zero
value $X_{0}$ in the physical Higgs vacuum of the theory. The value
which goes into \eqref{eq:gauginomu1} is thus \begin{equation}
\frac{F^{X}}{X_{0}}=\frac{1}{2}\left(\frac{F^{u}}{v_{u}}+\frac{F^{d}}{v_{d}}\right),\label{eq:FX/X}\end{equation}
 where we have set the vevs of the charged components of the Higgs
fields to zero and $v_{u}$($v_{d}$) are the vevs of the neutral
Higgses $H_{u}^{0}$($H_{d}^{0}$). Now the F-terms may be computed
from (\ref{eq:Wmssm}) and \eqref{eq:Kfull1}. The relevant term in
the superpotential is $W\sim\mu H_{u}H_{d}$ and in the Kaehler potential
it is $K\sim Z(H_{u}\bar{H}_{u}+H_{d}\bar{H_{d})}$. Then we have
(with the indices for $H_{u},H_{d}\rightarrow u,d$)

\begin{eqnarray}
F^{\bar{u}} & = & e^{K/2}K^{\bar{u}u}(\partial_{u}W+K_{u}W)=e^{K/2}K^{\bar{u}u}\tilde{\mu}H_{d}+m_{3/2}\bar{H}_{u}\simeq m_{3/2}v_{u},\label{eq:Fu}\\
F^{\bar{d}} & = & e^{K/2}K^{\bar{dd}}(\partial_{d}W+K_{d}W)=e^{K/2}K^{\bar{d}d}\tilde{\mu}H_{u}+m_{3/2}\bar{H}_{d}\simeq m_{3/2}v_{u}.\label{eq:Fd}\end{eqnarray}
As is usual in the MSSM we have chosen the vevs to be real (this may
in fact be done without loss of generality) and we have ignored the
`mu' term contribution since it is suppressed in our class of models.
Thus we have \begin{equation}
\frac{F^{X}}{X_{0}}=m_{3/2}.\label{eq:FX/X0}\end{equation}
Using this and \eqref{eq:Bmugrana} in \eqref{eq:gauginomu1} we get
\begin{equation}
\frac{m_{a}}{g_{a}^{2}}(\mu)=-\frac{b_{a}}{8\pi^{2}}m_{3/2}.\label{eq:gauginomass2}\end{equation}
 The above discussion is in fact the usual treatment of RG evolution
that is used in the presence of thresholds. In our case this threshold
is at the soft mass scale which is in fact the same as the Higgs scale
$v$. Above this scale all superparticles would contribute to the
evolution, while below one might expect only the standard model particles
to contribute. This for instance is the assumption made in extrapolating
from the standard model to the GUT scale to get unification, by accounting
for the superpartners of the standard model particles which give a
similar threshold effect. Note that these masses are of the same order
of magnitude as the squark/slepton masses. For completeness we quote
the values obtained for each separate gaugino\begin{eqnarray}
m_{3} & = & -7\frac{\alpha_{3}}{4\pi}m_{3/2}\label{eq:gaugino3}\\
m_{2} & = & -\frac{19}{6}\frac{\alpha_{2}}{4\pi}m_{3/2}\label{eq:gaugino2}\\
m_{1} & = & \frac{41}{10}\frac{\alpha_{1}}{4\pi}m_{3/2}\label{eq:gaugino1}\end{eqnarray}

In the usual discussion of AMSB it is claimed that there is a contribution
from Weyl anomalies to the soft masses as well \citep{Randall:1998uk}\citep{Pomarol:1999ie}.
This argument is based on the following reasoning. One starts with
the \textit{assertion} that the wave function renormalization should
undergo the following replacement \begin{equation}
Z(\Phi,\bar{\Phi};\ln\frac{\Lambda}{\mu})\rightarrow Z(\Phi,\bar{\Phi};\ln\frac{\Lambda|C|}{\mu})\label{eq:AMSB?}\end{equation}
in supergravity. Then one has for the squared soft masses\begin{equation}
m^{2}=-\ln Z|_{\theta^{2}\bar{\theta}^{2}}=-\frac{1}{4}|F^{C}|^{2}\frac{d^{2}\ln Z}{d\ln\Lambda^{2}}.\label{eq:AMSBsoftmass?}\end{equation}
However not only is there no justification for the replacement \eqref{eq:AMSB?},
making $Z$ dependent on $C$ would in fact violate the Weyl gauge
invariance of supergravity which is essential for the consistency
of the formalism. In fact as discussed above, the addition of a $\ln C$
term to the gauge coupling function by KL \citep{Kaplunovsky:1994fg}
was designed to restore the Weyl invariance of the theory. The replacement
\eqref{eq:AMSB?} on the other hand would result in breaking the Weyl
invariance making $C$ a propagating field, and therefore it is incorrect.
Formula \eqref{eq:AMSBsoftmass?} is therefore invalid.

Nevertheless there is a contribution to the soft mass that comes from
a quantum effect that has nothing to do with Weyl anomalies. This
was pointed out in \citep{Dine:2007me} and the mechanism is a consequence
of the formula \eqref{eq:HRGX}. In the Higgs branch of the theory
The radiatively generated soft mass at a scale $\mu\rightarrow X_{0}$
\begin{equation}
m_{\Phi}^{2}(X_{0})=2\sum_{a}c^{a}(r)\left(\frac{\alpha_{X^{0}}^{a}}{4\pi}\right)^{2}(b^{a}-b^{'a})\frac{|F_{X}|^{2}}{|X_{0}|^{2}}\label{eq:m2X}\end{equation}
\footnote{Again this is very similar to the argument given in \citep{Giudice:1997ni}
for the corresponding GMSB calculation with similar caveats as in
the gaugino case.%
}. Here the sum is taken over the three gauge group factors and $\alpha^{a}=g^{a2}/4\pi$.
Also $c^{a}(r)$ is the quadratic Casimir of the gauge group representation
$r$ of the observable field (squark or slepton). Then using (\ref{eq:FX/X0})
we have the contribution to the soft masses,\begin{equation}
m_{\Phi}^{2}(X_{0})=2\sum_{a}c_{\Phi}^{a}\left(\frac{\alpha_{X^{0}}^{a}}{4\pi}\right)^{2}(b^{a}-b'^{a})m_{3/2}^{2}.\label{eq:m2X2}\end{equation}
As we argued earlier, above the scale $X_{0}$ all superparticles
would contribute to the evolution, while below one might expect only
the standard model particles to contribute. This gives $b_{3}-b'_{3}=4,\, b_{2}-b'_{2}=25/6,\, b_{1}-b'_{1}=5/2.$
We also have the values (with $Q,L$ standing for the quark, lepton
doublets respectively) $c_{Q}^{3.2.1}=\frac{4}{3},\frac{3}{4},\frac{1}{60}$;
$c_{L}^{2,1}=\frac{3}{4},\frac{3}{20}$; $c_{\tilde{u}}^{3,1}=\frac{4}{3},\frac{4}{15}$
and $c_{\tilde{d}}^{3.1}=\frac{4}{3},\frac{1}{15}$. The formula then
gives the following generation independent contribution to the masses
of the squark and sleptons. \begin{eqnarray}
m_{Q}^{2} & = & \left[\frac{32}{3}\left(\frac{\alpha_{3}}{4\pi}\right)^{2}+\frac{25}{4}\left(\frac{\alpha_{2}}{4\pi}\right)^{2}+\frac{1}{12}\left(\frac{\alpha_{1}}{4\pi}\right)^{2}\right]m_{3/2}^{2},\label{eq:Lsquark}\\
m_{\tilde{u}}^{2} & = & \left[\frac{32}{3}\left(\frac{\alpha_{3}}{4\pi}\right)^{2}+\frac{4}{3}\left(\frac{\alpha_{1}}{4\pi}\right)^{2}\right]m_{3/2}^{2},\label{eq:usquark}\\
m_{\tilde{d}}^{2} & = & \left[\frac{32}{3}\left(\frac{\alpha_{3}}{4\pi}\right)^{2}+\frac{1}{3}\left(\frac{\alpha_{1}}{4\pi}\right)^{2}\right]m_{3/2}^{2},\label{eq:dsquark}\\
m_{L}^{2} & = & \left[\frac{25}{4}\left(\frac{\alpha_{2}}{4\pi}\right)^{2}+\frac{3}{4}\left(\frac{\alpha_{1}}{4\pi}\right)^{2}\right]m_{3/2}^{2},\label{eq:slep}\\
m_{\tilde{e}}^{2} & = & 3\left(\frac{\alpha_{1}}{4\pi}\right)^{2}m_{3/2}^{2}.\label{eq:selectron}\end{eqnarray}
Let us now compare with the contribution from the quadratic divergence
effects with $h_{21}\sim3-5\times10^{2}$. For the squark masses the
contribution from \eqref{eq:msugraparam}, is somewhat smaller than
the values in \eqref{eq:Lsquark}-\eqref{eq:dsquark} but it is an
order of magnitude larger than the contribution \eqref{eq:slep}\eqref{eq:selectron}
to the slepton masses.

Note that the equations (\ref{eq:gaugino3}) to \eqref{eq:gaugino1}
and \eqref{eq:Lsquark} to \eqref{eq:selectron} give masses in the
$O(10^{3}GeV)$ to $O(10^{2}GeV)$ range, provided we choose, as in
the case of the usual presentation of AMSB phenomenology, a large
mass for the gravitino ($\sim30TeV$). Unlike that case however we
do not require an additional mechanism to get non-negative squared
slepton masses even though here we actually have such a contribution,
namely \eqref{eq:m2correction}, which as we observed above is much
larger than this `AMSB' contribution. Note that these give masses
in the $O(10^{3}GeV)$ to $O(10^{2}GeV)$ range with the above choice
for the gravitino mass.

\section{Summary of phenomenology and conclusions}

In this paper we have studied in detail the SUSY breaking phenomenology
of a classically `no-scale' like or `sequestered' model (in the sense
that the classical soft masses are suppressed relative to the gravitino
mass) based on type IIB string theory with the MSSM coming from open
string fluctuations on a stack of D3 branes. While there does not
yet exist a complete chiral theory of this sort (with all moduli stabilized)
it is plausible that once various technical difficulties are overcome
such a model can indeed be constructed. If that turns out to be the
case then its qualitative phenomenology would be that discussed in
this paper. Actually it should be clear from the arguments that we
have made, that this kind of phenomenology is quite generic for theories
which are of this type (i.e. with suppressed classically generated
soft terms). Thus we expect that similar phenomenological results
emerge from a large class of string theoretic SUGRA models such as
for instance IIB models with the MSSM on D7 branes.

These models have features that are similar to those found in all
three standard mechanisms of SUSY mediation.

\begin{itemize}
\item Origin of SUSY breaking in moduli and transmitted by gravity as in
mSUGRA. 
\item Soft parameters are due to quantum effects as in AMSB and GMSB.
\item Gaugino masses mainly from `AMSB'.
\item ${\color{black}m_{3/2}\gtrsim10TeV}$ as in AMSB.
\end{itemize}
Such models have just two parameters that can be adjusted - the gravitino
mass and the (integer) $h_{21}$ (or in more general models the sum
$h_{21}+h_{11}$). The cutoff is almost completely fixed once we demand
that it should be higher than the scale at which the gauge couplings
appear to unify, but below the string scale. Since at this point unification
is the only concrete (albeit rather tenuous) evidence for supersymmetric
physics, we strongly believe that it should be taken as an input.
Since the cutoff should be definitely less than the string/Planck
scale this limits us to the range $10^{16}GeV<\Lambda<10^{18}GeV$.
Furthermore as the string scale is expected to be somewhat below the
Planck scale (perhaps $\le10^{17}GeV$) we are actually restricted
to $\Lambda\sim10^{16}GeV=10^{-2}M_{P}$, which is the value that
we have used. Also as we have seen in section \ref{sec:Quadratic-divergence-issues}
the effective perturbative parameter is $h_{21}\Lambda^{2}/16\pi^{2}M_{P}^{2}\sim h_{21}10^{-6}$
with the above value of $\Lambda$. As we argued above, with a gravitino
mass of around $30TeV$, the number of cycles in the Calabi-Yau manifold
should not be much larger than $10^{2}$ since the $\mu$-term has
to be well below $1TeV$ (see equation \eqref{eq:Bmugrana}) . On
the other hand we cannot lower the gravitino mass below about $10TeV$
since in that case the gaugino masses (for the $SU(2)\times U(1)$
group) would be too low. Thus we see that this class of models must
have \begin{equation}
h_{21}\gtrsim3-5\times10^{2},\, m_{3/2}\sim10-30TeV.\label{eq:h21m3/2}\end{equation}
 The universal scalar masses and the $\mu$ and $B\mu$ terms (at
the unification scale $\Lambda$) are \begin{equation}
\mu\sim\frac{B\mu}{\mu}\sim m_{s}\sim100-500GeV,\label{eq:muBmums}\end{equation}
and the gaugino masses (which are naturally computed at the MSSM scale
via the `AMSB' calculation of \citep{Dine:2007me,deAlwis:2008aq}
(see equations \eqref{eq:gaugino3}\eqref{eq:gaugino2}\eqref{eq:gaugino1})
are \[
m_{1}\sim30-80GeV,\, m_{2}\sim40-100GeV,\, m_{3}\sim400-1000GeV.\]

Finally we should stress that so far there is no concrete string theoretic
construction of the MSSM (living on D3 branes in type IIB or in any
other string theory set up) with all the moduli stabilized. In this
paper we have assumed that the SUSY breaking phenomenology of non-chiral
constructions that has been discussed in the literature \citep{Grana:2003ek}\citep{Camara:2004jj}
will hold for chiral models as well. Of course as we discussed before
it is possible that these models will have FCNC terms that are proportional
to $h_{21}$, like the flavor conserving terms, and in this case one
would need a fine tuning of one part in $10^{3}$ to suppress them.

Even with such additional fine-tuning it seems that this class of
models is still the minimal possible and least fine-tuned one that
can be embedded in string theory. Suppose for example there is a mSUGRA
model with all moduli stabilized which does not have FCNC at the classical
level.  One would still have FCNC terms, but since the classical contribution
to the scalar squared mass is now $O(m_{3/2}^{2})$ the quantum contribution
will be down by a factor $h_{21}\frac{\Lambda^{2}}{16\pi^{2}}\sim10^{-4}$
and can be ignored. However now the gravitino mass is low $\sim100GeV$
so there is a fine-tuning factor (according to the work of Douglas
and collaborators \citep{Douglas:2004qg}) $O(\frac{m_{3/2}(low)}{m_{3/2}(high)})^{6}=O(10^{-12})$
when the high value is taken to be $\sim10TeV$ as in the model discussed
here. As for GMSB models, one might expect that the same factor applies,
however it has been argued that such models are on a different branch
\citep{Dine:2004is}. Be that as it may, GMSB requires an additional
sector - the so-called messenger sector - compared to the class of
models discussed here.

It is clearly important to find detailed constructions that incorporate
the MSSM within a string theory context where all moduli can be stabilized
\footnote{For recent progress on such constructions see \citep{Aparicio:2008wh}
\citep{Beasley:2008dc,Beasley:2008kw}.%
}. Even though one may not be able to make precise predictions (since
by changing the fluxes one can make changes to the masses and couplings)
the physics that we have discussed above would then be a qualitative
prediction of string theory for LHC physics. This is so in the sense
that the class of models that we have here are the minimal possible
in terms of fine-tuning, and having just the sectors (namely a visible
MSSM or GUT sector and a moduli sector) that necessarily have to be
present.

\section{Acknowledgments}

I wish to acknowledge discussions with Oliver DeWolfe. I also wish
to thank Yuri Shirman for discussions and useful comments on the manuscript.
This research is supported in part by the United States Department
of Energy under grant DE-FG02-91-ER-40672. Finally I would like to
thank the Aspen Center for Physics where the revisions to this paper
were made.

\section*{Appendix - On the Relation to Mirage Mediation and Large Volume Scenarios}

The phenomenological consequences of the class of models discussed
in this paper have a certain superficial resemblence to the so-called
mirage mediation models of \citep{Choi:2005ge} (see also \citep{Everett:2008qy}
for a recent variant of this) where the classical (plus nonperturbative)
contribution is of the same order as the `AMSB' one. However these
models rely on the KKLT toy model with an uplift term. There are several
problems with taking this model seriously for phenomenological purposes.
Firstly (as pointed out in \citep{Brustein:2004xn}) the logic of
deriving a four dimensional theory from a ten dimensional theory requires
that one starts from a classical supersymmetric vacuum of the ten-dimensional
theory which enables one to organize the fluctuations around that
point in 4D supergravity multiplets, and will then necessarily give
a 4D ${\cal N}=1$ SUGRA theory. If one starts with supersymmetry
broken at the string level (even if the scale is suppressed by warping)
there is simply no way of deriving a four dimensional supergravity.
In fact the potential one gets is a runaway one for the Kaehler moduli
- implying decompactification. The potential that is usually used
is based on the assumption that one can add a non-perturbative term
to the superpotential before one adds the Dbar term - but this is
an inversion of the logic since the string theory starting point did
not have such a non-perturbative term to begin with (especially if
they arise from low energy gauge theoretic effects) while the Dbar
brane is added at the string theory level i.e. in the ten dimensional
theory.

If we ignore this, the uplift term is an explicit breaking of the
${\cal N}=1$ four-dimensional supersymmetry (although in 10 dimensions
it is a spontaneaous breaking caused by Dbar branes) at an intermediate
scale $\sqrt{m_{3/2}M_{P}}\sim10^{11}GeV$. The Dbar brane is located
far down a throat so that its effective tension and hence the supersymmetry
breaking, is warped down from the string/Planck scale to the above
scale by the warp factor $e^{A_{min}}\sim\sqrt{m_{3/2}/M_{P}}$ (see
\citep{Choi:2005ge} equation (19)). If the MSSM/GUT branes are in
the bulk (as is effectively the case in \citep{Choi:2005ge}) as opposed
to being in the infra-red end of a throat region, this appears to
give quantum effects that result in terms $O(m_{3/2}M_{P}\Lambda^{2}/16\pi^{2})$
(rather than $O(m_{3/2}^{2}\Lambda^{2}/16\pi^{2})$ as in this paper)
in the potential. This would seem to introduce large corrections to
soft masses etc when $\Lambda\sim M_{GUT}$. However the overlap of
the wave function of the MSSM states which are located at the UV end
of the throat with the SUSY breaking fields at the IR end of the throat
is exponentially suppressed. This results in an effective mass splitting
at the UV end $\Delta m^{2}\sim e^{2A_{min}}m_{3/2}M_{P}=m_{3/2}^{2}$.
and hence the quadratically divergent quantum contribution is again
of the same order as in this paper. Another possibility would be to
have the MSSM brane far down a throat region so that the effective
cut off $\Lambda$ is also warped down to an intermediate scale $\Lambda_{eff}^{2}=\frac{m_{3/2}}{M_{P}}\Lambda^{2}$,
so that the estimate of the quantum contribution to the potential
is again $O(m_{3/2}M_{P}\Lambda_{eff}^{2}/16\pi^{2})=O(m_{3/2}^{2}\Lambda^{2}/16\pi^{2})$.
In this case also the corrections to soft masses etc. will not significantly
change the classical contributions. However in such a situation it
is not entirely clear how to obtain a GUT theory since the effective
cut off is far below the GUT scale though in the corresponding 5-dimensional
scenario a possible resolution has been offered in \citep{Randall:2001gb}.
Also in the string theoretic case the derivation of the effective
four dimensional supergravity in the presence of significant warping
has not yet been entirely resolved (for the problems associated with
this and progress towards a resolution of this question see \citep{Burgess:2006mn}\citep{Shiu:2008ry}).
It should also be mentioned that the mirage mediation scenario is
not necessarily tied to having an anti-brane at the IR end of the
throat. This could in principle be replaced by a conventional SUSY
breaking sector at the IR end of a warped throat (for some discussion
of this see \citep{Brummer:2006dg})%
\footnote{I wish to thank an anonymous referee for correcting the estimate of
the quantum effect in the Mirage Mediation case made in an earlier
version of this paper and for drawing my attention to some  relevant
references.%
}.

One might of course avoid large corrections by taking the cutoff $\Lambda$
to be much smaller than the GUT scale and this appears to be the case
in the large volume scenario (LVS) discussed in \citep{Balasubramanian:2005zx}.
In that and in subsequent work based on it, it is shown that in the
absence of fine tuning a large volume scenario emerges (from GKP-KKLT
type constructions) where the string scale is an intermediate scale
(around $10^{12}GeV$). In this case there is a broken SUSY minimum
with negative CC (exactly as required for the classical construction
of this paper) and the authors use the uplift term of KKLT to lift
the minimum to a small positive value. However the phenomenology appears
to be insensitive to the uplift term (see for example \citep{AbdusSalam:2007pm}).
But as we've argued in this paper, at this point in time, supersymmetric
grand unification is  the main piece of evidence for supersymmetry
and should be taken seriously in model building. This can be achieved
in the LVS scenario (if one fine-tunes the flux generated superpotential
as we've done in this work)  but then the quadratically divergent
corrections that we have discussed in this paper will become relevant.
A detailed discussion of this as well as an extention of the phenomenology
discussed in this paper to the case when the standard model is located
on D7 branes, will appear in future work.

\bibliographystyle{apsrev} \bibliographystyle{apsrev}
\bibliography{myrefs}

\end{document}